\documentclass[
	a4paper, 
	10pt, 
	twoside, 
]{LTJournalArticle}
\usepackage{amsthm,amsmath,amssymb,bbm,bm}
\DeclareUnicodeCharacter{0301}{*************************************}

\title{An Article Title That Spans Multiple\\ Lines to Show Line Wrapping} 

\author{Yuhan Li\textsuperscript{1$\ast$},  Hongtao Zhang\textsuperscript{2}\thanks{The first and second authors contribute equally.}, Keaven Anderson\textsuperscript{2}, Songzi Li\textsuperscript{3} and Ruoqing Zhu\textsuperscript{1}\thanks{Corresponding author: \href{mailto:rqzhu@illinois.edu}{rqzhu@illinois.edu}
}}

\date{\footnotesize\textsuperscript{\textbf{1}} Department of Statistics, University of Illinois Urbana-Champaign, Champaign, IL, USA\\ 
\textsuperscript{\textbf{2}} Biostatistics and Desicion Research Science, Merck \& Co., Inc., North Wales, PA, USA\\
\textsuperscript{\textbf{3}} Biostatistics, Agenus Inc., Lexington, MA, USA}

\usepackage{setspace}
\begin{document}
\title{AI in Pharma for Personalized Sequential Decision-Making: Methods, Applications and Opportunities}
\maketitle
\setcounter{page}{1}
\pagenumbering{arabic}

\vspace{-3mm}
\section{Introduction}
\vspace{-3mm}
In the pharmaceutical industry, the use of artificial intelligence (AI) has seen consistent growth over the past decade. This rise is attributed to major advancements in statistical machine learning methodologies, computational capabilities and the increased availability of large datasets. AI techniques are applied throughout different stages of drug development, ranging from drug discovery to post-marketing benefit-risk assessment. Kolluri et al. \cite{kolluri2022machine} provided a review of several case studies that span these stages, featuring key applications such as protein structure prediction, success probability estimation, subgroup identification, and AI-assisted clinical trial monitoring. From a regulatory standpoint \cite{liu2023landscape}, there was a notable uptick in submissions incorporating AI components in 2021. The most prevalent therapeutic areas leveraging AI were oncology (27\%), psychiatry (15\%), gastroenterology (12\%), and neurology (11\%).

The paradigm of personalized or precision medicine has gained significant traction in recent research, partly due to advancements in AI techniques \cite{hamburg2010path}. This shift has had a transformative impact on the pharmaceutical industry. Departing from the traditional ``one-size-fits-all'' model, personalized medicine incorporates various individual factors, such as environmental conditions, lifestyle choices, and health histories, to formulate customized treatment plans. By utilizing sophisticated machine learning algorithms, clinicians and researchers are better equipped to make informed decisions in areas such as disease prevention, diagnosis, and treatment selection, thereby optimizing health outcomes for each individual \cite{chakraborty2013statistical,kosorok2019precision}.

In this article, we explore a range of methods and algorithms in the field of personalized medicine. While these techniques share the overarching aim of crafting personalized treatment plans, they differ in terms of problem formulations and practical applications. We delve into specific examples within the healthcare sector, categorizing them as either established in research and practice, or as aspirational approaches with potential for significant impact. The article concludes with a discussion of pertinent challenges and outlines avenues for future research.

\clearpage

\section{Methods and Applications}
\vspace{-5mm}
\subsection{Optimal Treatment Sequence}
\label{finite}
\vspace{-2mm}

\textbf{Dynamic Treatment Regime}\, Dynamic Treatment Regime (DTR) represents a cutting-edge paradigm in the realm of personalized medicine, aiming to tailor medical interventions to individual patients' evolving health status \cite{laber2014dynamic}. Within the context of clinical research, data concerning a DTR are usually collected from multi-stage clinical trials or longitudinal observational studies on the disease of interest \cite{clifton2020q}. These studies often involve a finite number of decision stages. An optimal DTR aims to find a sequence of decision rules that assign treatments at each stage based on a patient's baseline characteristics and historical information.

Suppose we have a pre-specified finite \(T\) decision points, indexed by \(t=1,2,\ldots,T\). Let \(S^t \in \mathbb{R}^p\) represent all related patient characteristics at time \(t\), such as age, gender, and lab results, which may vary across time points and reflect the patient's current condition. The treatment given at time \(t\) is denoted by \(A^t \in \mathcal{A}\), which may include drug choice and/or dosage selected from a possible set of treatments \(\mathcal{A}\), which could be either discrete or continuous. The potential treatment trajectory up to point \(t\) is denoted by \(\bar A^t = (A^1,A^2,\ldots,A^t)\), and \(\bar S^t = (S^1,S^2,\ldots,S^t)\) represents the cumulative information of the patient leading up to \(t\). Realizations of such a treatment path and accumulated patient data are denoted as \(\bar a^t = (a^1,a^2,\ldots,a^t)\) and \(\bar s^t = (s^1,s^2,\ldots,s^t)\), respectively.

At each decision point, we further observe an immediate reward \(R^t\) that may depend on all the previous history leading up to \(t\). The immediate reward serves as an indicator of the individual's response to the selected treatment, where a larger reward signifies a more favorable response. Hence, the collected dataset is of the form  \(\{S_i^1,A_i^1,R_i^1,S_i^2, \ldots ,S_i^T,A_i^T,R_i^T,S_i^{T+1}\}^n_{i=1}\), which comprises \(n\) i.i.d. trajectories with \(T\) decision points. The objective is to identify the optimal DTR that maximizes the cumulative reward, i.e., \(R = \sum^T_{t=1}R^t\), from \(t=1\) to \(T\). Under certain scenarios, it is also possible to only observe the final reward \(R\) at the last stage, e.g., event-free survival or overall survival. In either case, the goal is to maximize \(R\) by choosing a sequence of decisions.

A DTR is defined as \(\boldsymbol \pi = (\pi_1,\ldots,\pi_T)\), which forms a sequence of decision rules to treat a patient over time. The decision rule at each time point \(t\), \(\pi_t\), can be thought of as a mapping from a patient's history \(\bar S^t\) to the available treatment option set \(\mathcal{A}\). The optimal DTR \(\boldsymbol {\pi^*} = (\pi_1^*,\pi_2^*,\ldots,\pi_T^*)\) is defined as the DTR that achieves the maximum expected reward, i.e., \(R_{\boldsymbol \pi} = \mathbb{E}(\sum^T_{t=1}R^t_{\pi_t}) \leq \mathbb{E} (\sum^T_{t=1}R^t_{\pi_t^*}) = R_{\boldsymbol \pi^\ast}\) for all \(\boldsymbol \pi\).

\noindent \textbf{Q-learning} \, To estimate the optimal DTR, Q-learning is widely used, particularly in the finite-horizon setting where decision stages are limited and predetermined \cite{murphy2005generalization, robins2004optimal, zhao2009reinforcement, nahum2012q, kosorok2015adaptive}. Q-learning adopts a backward induction mechanism, starting its estimation at the last decision point and working its way back to the beginning.  


We begin at the final stage \( T \) and posit models such as linear models or random forests to estimate the \( Q_T(\bar{s}^T, \bar{a}^{T}) \) \cite{zhao2009reinforcement}. The observed cumulative reward \( \sum_{t=1}^{T} R^t \) serves as the response variable, while \( \bar{s}^T \) and \( \bar{a}^{T-1} \) are used as covariates. Once the model is estimated, we identify the treatment \( \hat{\pi}^*_{T} \) that maximizes the expected reward for a patient at decision point \( T \), given their historical profile \( \bar{s}^T, \bar{a}^{T-1} \). To find the optimal treatment regime, we work backward, and treat the already-estimated Q-function value as the new response variable for previous decision points. Following similar steps, we estimate \( \boldsymbol{\hat{\pi}^*} = (\hat{\pi}_1^*, \ldots, \hat{\pi}_T^*) \) as the overall optimal treatment regime. For a more comprehensive discussion, we refer readers to \cite{clifton2020q}.


In a finite-horizon setting, Q-learning has gained tremendous popularity due to its ease of implementation and overall strong performance \cite{kosorok2019precision}. Nonetheless, its sensitivity to model misspecification presents a challenge. If the posited model for Q-functions is misspecified, performance may suffer significantly, as the bias would backpropagate to the very first stage. Various alternatives and variations have been proposed to address such limitations. For instance, Advantage Learning (A-learning, \cite{murphy2003optimal}) estimates the optimal DTR by modeling the difference in outcomes between two treatment options, making it more robust to model misspecification \cite{schulte2014q}. Robust Q-learning \cite{ertefaie2021robust} introduces data-adaptive techniques for nuisance parameter estimation, tackling both residual confounding and efficiency loss.

There are also other notable advancements and extensions in Q-learning, such as statistical inference for Q-learning based on the asymptotic normality of estimators \cite{song2015penalized} and bootstrap methods \cite{chakraborty2014inference}. \cite{murray2018bayesian} developed a Bayesian framework for finding the optimal DTR to accommodate prior knowledge and measure the uncertainty of the estimated DTR. \cite{cho2023multi} extended original Q-learning methods to survival outcomes, and \cite{zhu2019proper} considers high-dimensional settings and variable selection in Q-learning. For an exhaustive overview, we direct readers to \cite{kosorok2019precision, clifton2020q}.

\noindent \textbf{Application: Treatment Regime in Perioperative Setting}\, Numerous studies have explored the application of Q-learning and its variants in clinical trial settings, aiming to find the optimal DTR from clinical trial data \cite{zhao2009reinforcement,zhao2011reinforcement,yauney2018reinforcement}. To illustrate this, we consider the treatment of early-stage malignant tumors, which could be surgically removed at this stage. The perioperative process typically begins with neoadjuvant therapy, designed to shrink the tumor and thereby enhance the chances of a successful surgery. Following the operation, adjuvant therapy is administered to prevent cancer recurrence.

We can model this as a Q-learning problem with two decision points. The state variables might include factors such as tumor stage, resection margin (R0/R1/R2), pathology, tumor imaging data, and patient health status. The first decision action, \(A^1\), is the choice of neoadjuvant treatment. The second decision action, \(A^2\), could be two-dimensional, incorporating both the choice of adjuvant treatment and its duration (number of cycles). Event-free survival (EFS) can serve as the final reward. By integrating the corresponding covariates into the Q-learning framework, we can estimate the optimal treatment sequence for both the neoadjuvant and adjuvant periods, tailored to the characteristics of individual patients.


\noindent \textbf{Application: Lines of Therapies for Metastatic Cancers}\, When treating metastatic cancer, the typical medical practice is to treat patients with the same drug until either the disease progresses or the patient becomes intolerant to the drug. The next-in-line treatment is then initiated. Identifying a personalized optimal treatment regime or sequence with the aim of maximizing a certain metric, such as overall survival, is of tremendous significance in healthcare. Given the multiple decision points associated with prescribing next-in-line treatments, such as drug choice and the time point to switch to the next-line treatment, Q-learning becomes a natural fit for tackling this issue. For example, \textcite{zhao2011reinforcement} illustrated their reinforcement learning (RL) model in the context of lines of chemotherapy for metastatic non-small-cell lung cancer (NSCLC).

Fast-forward to the era of immuno-oncology, the narrative has evolved to focus on the personalized optimal sequence involving the PD-1/PD-L1 checkpoint inhibitors: Should they be given as monotherapy or in combination with other drugs, in what order, and to which patients? Most major checkpoint inhibitors, such as pembrolizumab and nivolumab, have been evaluated either as monotherapy or in combinations in different lines to treat patients with metastatic NSCLC. Therefore, existing data may already hold the answers to these questions. Applying Q-learning and other appropriate RL methods to the aggregated data could provide extremely valuable insights for improving the treatment of these patients.


\vspace{-5mm}
\subsection{Adaptive Clinical Trial Design}
\vspace{-2mm}
\textbf{Adaptive Clinical Trials}\, Q-learning is primarily concerned with estimating optimal DTR using pre-collected datasets. However, adaptive clinical trials require real-time, data-dependent decision making, such as selecting treatment arms based on historical data up to a certain cutoff point \cite{zame2020machine}. This real-time utilization of cumulative data is known as the ``online setting'', which stands in contrast to the ``offline setting'' in which pre-collected datasets are used \cite{coronato2020reinforcement}.

To formalize this problem in the context of adaptive clinical trial design, we consider a trial with \( N \) treatment arms. Each arm \( i \) is associated with an unknown probability distribution \( D_i \), which describes the treatment outcomes (efficacy or toxicity) when assigning that particular treatment to a patient. At each decision point \( t \), a reward \( R^t \) is obtained from the corresponding distribution \( D_i \) when treatment arm \( i \) is selected. The objective is to determine the recommendation rule at each decision point based on the accumulated data. This rule aims to maximize the expected cumulative reward \( \mathbb{E} [ \sum_{t=1}^{T} R^t ] \).

Such formulation transforms the adaptive design into a multi-armed bandit (MAB) problem \cite{press2009bandit, villar2015multi}. The major challenge in solving such a problem lies in balancing the trade-off between ``exploration'', where less-understood arms are chosen to collect more data about their distributions, and ``exploitation'', where arms with higher observed cumulative rewards are chosen to maximize the expected outcome \cite{audibert2009exploration}. Therefore, effective solutions to the MAB problem in the context of adaptive clinical trials must address this exploration-exploitation dilemma to achieve optimal patient outcomes.

\noindent \textbf{Multiarmed Bandit}\, Various methods have been developed to tackle the MAB, such as the \( \epsilon \)-greedy algorithm \cite{lattimore2020bandit}, Thompson sampling \cite{agrawal2012analysis}, and Upper Confidence Bound \cite{garivier2011upper}, among others. The \( \epsilon \)-greedy algorithm takes a straightforward approach to the exploration-exploitation dilemma. With probability \( 1 - \epsilon \), the algorithm selects the arm with the highest empirical mean reward observed so far, known as the ``greedy'' action. With probability \( \epsilon \), it randomly selects an arm, thereby exploring the action space. The parameter \( \epsilon \) controls the trade-off between exploration and exploitation. A higher \( \epsilon \) promotes more exploration at the cost of immediate reward, while a lower \( \epsilon \) focuses more on exploitation. Meanwhile, Thompson sampling takes a Bayesian approach to the MAB problem. It maintains a probability distribution over the expected reward for each arm, updating these distributions as more data are collected. At each round \( t \), a sample is drawn from each arm's posterior distribution, and the arm with the highest sample is selected. The Upper Confidence Bound (UCB) algorithm selects the arm with the highest upper confidence bound on its expected reward. At each time step, it calculates the upper bound for each arm using both the estimated mean reward and its uncertainty. The arm with the highest calculated upper bound is then selected, aiming to minimize long-term regret.

\( \epsilon \)-greedy is straightforward and computationally efficient but suffers from constant, often unnecessary, exploration due to its \( \epsilon \) parameter \cite{white2013bandit}. Thompson sampling provides a more nuanced balance between exploration and exploitation by incorporating uncertainty through probabilistic models \cite{agrawal2012analysis}. While this leads to better performance in complex environments, it may require greater computational resources, particularly for complex posterior distributions \cite{umami2021comparing}. UCB has strong theoretical bounds on regret and is deterministic. However, it makes strong assumptions about the reward and can be less effective in non-stationary environments \cite{lattimore2020bandit}.

Several extensions to the original MAB algorithms have also been proposed to address real-world challenges, such as the analysis on sample complexity of MAB \cite{mannor2004sample}, MAB under dependent arms \cite{pandey2007multi}, MAB with safety constraints \cite{daulton2019thompson,shen2020learning}, and MAB with multiple objectives \cite{yahyaa2015thompson}. To further incorporate the patient-specific information to the decision-making process, contextual bandit framework has been introduced with additional state variables \cite{tewari2017ads,varatharajah2022contextual}. Such extension enables personalized treatment recommendations in adaptive clinical trials.

In the pharmaceutical setting, the MAB framework has been employed to study oncology dose-finding and response-adaptive randomization designs. We elaborate the first application and refer the readers to \cite{villar2015multi} for the latter. 

\noindent \textbf{Application: Oncology Dose-Finding}\, One primary objective of phase I oncology dose-finding trials is to identify the maximum tolerated dose (MTD) of the drug candidate to inform the dose level(s) to be investigated in subsequent phases of development. They start treating one cohort of patients, usually of size 3, at the lowest provisional dose level. Upon observing the data of the cohort, a recommendation (escalation/stay/de-escalation) is rendered regarding the dose level at which the next cohort of patients should be treated according to a certain statistical design. This process is repeated until the total sample size is exhausted or certain pre-specified early-stopping rules are met. 

Dose-finding has been an active area of statistical innovation. One important class of designs is the model-based designs \cite{OQuigleyEtal_1990, NeuenschwanderEtal_2008BLRM, zhang2022improving}. These designs postulate a parametric form of the dose-toxicity relationship and utilize the cumulative data to make a dose recommendation. The endpoint in most cases is a binary indicator of the presence of dose-limiting toxicity (DLT) within a certain period (e.g., 28 days). Patient-level covariate information can be intuitively incorporated in the model-based designs \cite{Neuenschwander_2015BLRMCombination}. 

Dose-finding trials are great candidates for applying the MAB framework due to their sequential and adaptive nature \cite{aziz2021multi, jin2023multiarmed}. Specifically, patients in the \( t^{th} \) cohort are assigned to dose \( D_t \) from the set of provisional doses \( \{1, \ldots, K\} \). The objective is to identify the dose level that is closest to the pre-specified target toxicity rate \( \theta \). Mathematically, this can be expressed as \( k^* = \arg \min_k |\theta - p_k| \), where \( p_k \) is the observed toxicity rate at dose $k$. We define the reward function \( R^t \) as \( R^t = - |\theta - \hat{p}^t_{D_t}| \), where \( \hat{p}^t_{D_t} \) is the estimated toxicity rate of the selected dose for cohort \( t \). By employing suitable MAB algorithms, the optimal dose level can be effectively identified. 

In recent years, the need for precision medicine is emerging more and more frequently with the development of new cancer treatments like T-cell engagers and cell therapies. Dosing of such therapies might need to be more personalized to avoid adverse events known to the mechanism of action, such as the cytokine release syndrome. The contextual bandit framework can be useful to incorporate patient-level information in this case \cite{varatharajah2022contextual}.




\vspace{-5mm}
\subsection{Mobile Health for Enhanced Patient Management}
\vspace{-2mm}
\noindent \textbf{Mobile Health (mHealth)} \, Section \ref{finite} details statistical methods for estimating optimal DTRs with a finite number of decision points. However, with the recent advancement of sensor technologies and wearable devices, it has become possible to record personal health information over an extremely long period with the help of mHealth technologies \cite{silva2015mobile}. Consequently, leveraging such data to formulate personalized treatment plans, addressing chronic diseases and various health issues across an infinite horizon with numerous decision points, has emerged as a prominent research area in recent years.

To date, mHealth has been used extensively in managing various health-related conditions including stress, depression, and other chronic diseases such as diabetes and cardiovascular diseases. It enhances patient monitoring and treatment for healthcare providers \cite{rehg2017mobile}. In mHealth settings, the data follows a similar pattern as Section \ref{finite}, which also consists of \( n \) i.i.d trajectories with \( T \) decision points, in the form of \( \{S_i^1,A_i^1,R_i^1,S_i^2, \ldots ,S_i^{T},A_i^{T},R_i^{T},S_i^{T+1}\}^n_{i=1} \). Compared with the finite horizon, several key differences should be noted.



First, the Markov property is assumed under the infinite horizon, meaning the next state and reward depend only on the current state and action, i.e., \( P(S^{t+1} = s^{t+1}| \bar S^t = \bar s^t,\bar A^t=\bar a^t)=P(S^{t+1} = s^{t+1}|S^t = s^t, A^t = a^t) \). Following the Markov property, the policy \( \pi \) is a function of the current state only, mapping it to a distribution on the action space where \( \pi(s)={P}(A^t=a|S^t=s) \). Finally, a discount factor \( \gamma \in [0,1) \) is introduced to ensure that the sum of rewards \( \sum^{\infty}_{k=0}\gamma^{k}R^{t+k} \) remains finite. A larger \( \gamma \) would place more weight on future rewards.

We generally model the whole process as a Markov decision process (MDP). An MDP is defined as a tuple \( <\mathcal{S},\mathcal{A},\mathbf{P},R,\gamma> \), where \( \mathcal{S} \) is the state space, \( \mathcal{A} \) is the action space, \( \mathbf{P}:\mathcal{S}\times \mathcal{A} \to \Delta(\mathcal{S}) \) is the unknown transitional kernel, \( R: \mathcal{S}\times \mathcal{S} \times \mathcal{A} \to \mathbb{R} \) is a bounded reward function, and \( \gamma \in [0,1) \) is the discount factor. A policy \( \pi \) is a mapping from the state space to the action space \( \pi: \mathcal{S}\to \mathcal{A} \). The goal is to find an optimal policy \( \pi^* \) that maximizes the expected discounted sum of rewards \( \mathbb{E}_{\pi}[\sum^{\infty}_{k=1}\gamma^{k-1}R^{t+k}|S^t=s] \).

\noindent \textbf{Reinforcement Learning (RL)} \, When the number of decision points approaches infinity, the task of determining the optimal policy transforms into a reinforcement learning (RL) problem \cite{sutton2018reinforcement}. In RL literature, we define the value function and state-value function for a given policy \( \pi \) as \( V_t^{\pi}(s)=\mathbb{E}[\sum^{\infty}_{k=0}\gamma^{k}R^{t+k}|S^t=s] \), and \( Q^{\pi}_t(s,a)=\mathbb{E}[\sum^{\infty}_{k=0}\gamma^{k}R^{t+k}|S^t=s,A^t=a] \). The only difference between the \( V \)-function and \( Q \)-function is whether we specify the action at time \( t \).

Based on these definitions, we can treat both the \( V \)-function and \( Q \)-function as measures of how good a policy is for a patient in any given state. By finding a policy that maximizes these quantities, we essentially achieve the goal of constructing a personalized treatment plan. However, this is not trivial given the dynamics over a long period, which can be difficult to model. Hence, the \emph{Bellman optimality equation} becomes an important tool.



We first define the optimal value function as \( V^*(s) = \max_{\pi} V^{\pi}(s) \), and the optimal $Q$-function is similarly defined as \( Q^*(s,a) = \max_{\pi} Q^{\pi}(s,a) \). These functions are interrelated through the equation \( V^*(s) = \max_{a} Q^*(s,a) \). The policy \( \pi^* \) that maximizes these functions is referred to as the optimal policy, denoted by \( V^{\pi^*}(s) = V^*(s) \) and \( Q^{\pi^*}(s, a) = Q^*(s,a) \). Both \( V^*(s) \) and \( Q^*(s,a) \) are unique and must satisfy the corresponding \emph{Bellman optimality equation} \cite{puterman2014markov}:
\vspace{-2mm}
{
\footnotesize
\begin{align*}
V^*(s) &= \max_{a} \mathbb{E}_{S^{t+1}|s,a} \left[ R^t + \gamma V^*(S^{t+1}) | S^t = s, A^t = a \right], \\
Q^*(s,a) &= \mathbb{E}_{S^{t+1}|s,a} \big[ R^t + \gamma \max_{a'} Q^*(S^{t+1}, a') | S^t = s, A^t = a \big].
\vspace{-4mm}
\end{align*}
}
\noindent Thus, \( V^*(s) \) and \( Q^*(s,a) \) serve as the fixed points of their respective Bellman optimality equations, and \( \pi^* \) can be solved accordingly.

One major challenge in solving the Bellman optimality equation arises when the dataset is collected under a policy that diverges from the optimal policy \( \pi^* \), while the Bellman optimality equation requires that actions be generated based on \( \pi^* \) to be valid \cite{fujimoto2019off}. Such distribution mismatch is the dominant case in mHealth setting and introduces both theoretical and computational challenges in finding the optimal policy.

To tackle these challenges, Greedy Gradient Q-learning (GGQ) \cite{ertefaie2018constructing} and V-learning \cite{luckett2019estimating} have been developed, formulating estimation equations based on the Q-function and V-function, respectively. GGQ has the advantage of enabling the construction of confidence intervals for the mean outcome difference between the optimal policy and any alternative policies. However, its estimation equation contains a non-smooth \(\text{max}\) operator, making estimation difficult without large amounts of data \cite{laber2014dynamic}. Furthermore, GGQ consistently selects the best arm at each decision stage, often resulting in sub-optimal outcomes in complex dynamic environments \cite{dann2014policy}. In contrast, V-learning adopts a stochastic policy distribution and avoids the non-smooth \(\text{max}\) operator, leading to more stable optimization. The stochastic policy class also makes V-learning more robust in the face of unexpected situations \cite{zhou2022estimating}.

While V-learning's stochastic policy class offers flexibility in action selection, it can degenerate into a uniform distribution in a large action space. To mitigate this, pT-learning was introduced, confining the support set to near-optimal actions at each decision point and allowing sparsity control through a tuning parameter \cite{zhou2022estimating}. Extending this, the Quasi-optimal Learning framework adapts the method to continuous action spaces, making it applicable to challenges such as optimal dose-finding over an infinite horizon \cite{li2022quasi}.

\begin{figure}[t]
\vspace{-2mm}
\centering
\includegraphics[width=8cm]{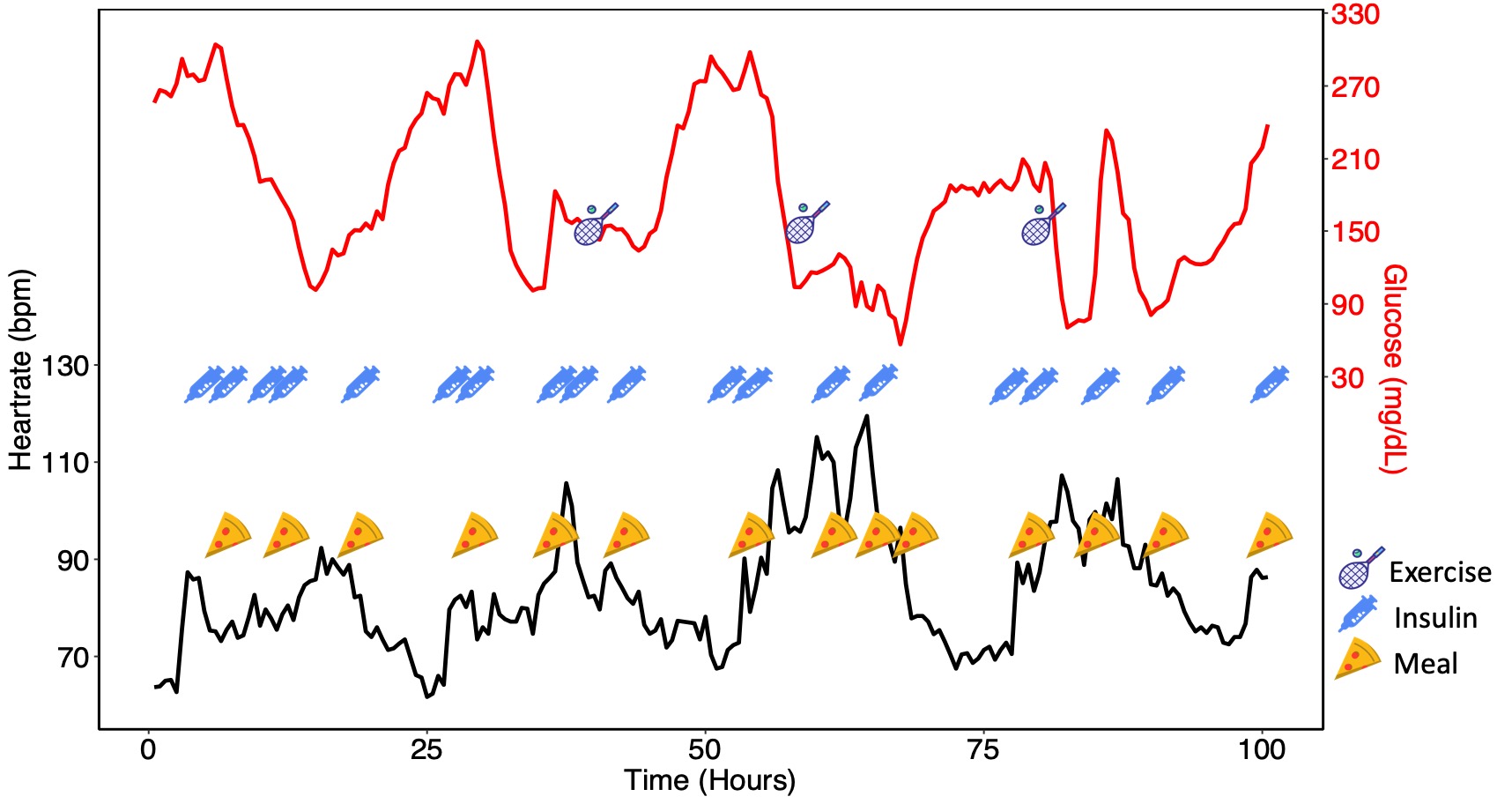}
\caption{OhioT1DM Data: A longitudinal observation of a patient}\label{fig:OhioT1DM}
\vspace{-4mm}
\end{figure}

\noindent \textbf{Application: Glucose Management for Diabetes}\, Glucose management in diabetes is a key mHealth application.  By continuously monitoring the glucose level, food intake, and physiological information, a series of just-in-time interventions, such as insulin injection, can be delivered to patients to improve long-term health outcomes \cite{muralidharan2017mobile}.  An application example is the OhioT1DM study \cite{marling2020ohiot1dm}, featuring 12 Type 1 Diabetes patients with continuous glucose monitoring (CGM) data, self-reported activity logs such as meal intakes and sleep status, and insulin injection dosages and timing over eight weeks. Figure \ref{fig:OhioT1DM} provides a snapshot of the fluctuation of glucose level, insulin injections, meals, exercise, and heart rate of a patient during a 100 hour time interval. 


As glucose dynamics can vary significantly between individuals, clinicians aim to personalize insulin injection doses based on each patient's health status \cite{bao2011improving}. Our objective is to develop a personalized treatment policy that optimally controls glucose levels for each individual.

We define the state variables as health status measurements for individual patients, and the action space refers to the insulin injection dose levels at each decision point. The glycemic index serves as the reward function, measuring the proximity of glucose levels to the normal range \cite{rodbard2009interpretation}. By applying methods like V-learning, pT-learning, and Quasi-optimal learning, we can determine an optimal policy for controlling each patient's glucose levels. Implementation details are available in \cite{luckett2019estimating,zhou2022estimating,li2022quasi}.

\vspace{-6mm}
\section{Discussion}
\vspace{-4mm}

We have introduced a wide range of methods and algorithms in personalized medicine. Under a finite horizon, methods like Q-learning and its variants, as well as MAB algorithms, have matured considerably in finding optimal DTRs and guiding the design of clinical trials. Nevertheless, these finite-horizon models have underlying assumptions that could be further relaxed to enhance their applicability.

Confounding and causality are critical issues in policy learning. Current methods often assume a fully observable environment; however, the true policy may be influenced by unmeasured confounders such as genetic factors \cite{coronato2020reinforcement}. Incorporating recent advances in causal inference to address these unmeasured confounders \cite{miao2018confounding,cui2023semiparametric} has emerged as a promising research direction \cite{kallus2018confounding,wang2022blessing,shi2022minimax}.

In offline settings like Q-learning, where data is pre-collected and no online interaction with the environment occurs, algorithms may suffer from inadequate coverage of state-action pairs. This can lead to imprecise estimations of value functions \cite{fujimoto2019off, xie2021bellman}. Hence, the pessimism principle is advised to limit the learned policy from visiting poorly-covered states, ensuring safety and avoiding undesired behaviors \cite{jin2021pessimism,uehara2021pessimistic}. Balancing pessimism and policy optimality represents another interesting research avenue \cite{ghasemipour2022so}.

Furthermore, the performance of an estimated DTR is assessed by its value function. Thus, it's essential to quantify uncertainties and conduct statistical inferences related to the value function. This challenge is closely tied to an emerging field of research known as off-policy evaluation (OPE), which aims to evaluate the value of a certain policy based on data generated from a different policy \cite{uehara2022review}. Notably, constructing confidence intervals for these value functions \cite{thomas2015high, feng2021non} and evaluating the value disparity between a particular policy and the optimal one are also pivotal research questions \cite{shi2022statistical}.

Under infinite horizons, in addition to the challenges present in finite horizons, further issues emerge that require extensive investigation. For example, the Markov property is a fundamental assumption under an infinite horizon. In mHealth settings, however, outcomes may be influenced by decisions made before the immediately preceding time point. Developing methods to test the validity of the Markov property \cite{shi2020does}, and to address violations in the data-generating process, is an important extension to existing frameworks.

Lastly, survival data is common in mHealth applications. Such data often includes treatment and covariate information that may be censored in follow-up stages, complicating policy learning. Although recent advancements in optimal policy estimation have been made within the survival data framework \cite{cho2023multi,zhao2020constructing, xue2022multicategory}, adapting these approaches to an infinite horizon remains a challenge.



\printbibliography

\end{document}